\documentclass{preprint}

\hypersetup{
    pdftitle={Discrete dispersion scan setup for measuring few-cycle laser pulses in the mid-infrared},
    pdfauthor={Nils C. Geib}
}

\newcommand{\T}{\tilde{T}}

\begin{document}
\title{Discrete dispersion scan setup for measuring few-cycle laser pulses in the mid-infrared}

\author[1,*]{Nils C. Geib}
\author[2,3]{Richard Hollinger}
\author[4]{Elissa Haddad}
\author[2]{Paul Herrmann}
\author[4]{François Légaré}
\author[1,6]{Thomas Pertsch}
\author[2,3]{Christian Spielmann}
\author[2,5,6]{Michael Zürch}
\author[1,7]{Falk Eilenberger}

\affil[1]{Institute of Applied Physics, Abbe Center of Photonics, Friedrich Schiller University Jena, Albert-Einstein-Str. 15, 07745 Jena, Germany}
\affil[2]{Institute of Optics and Quantum Electronics, Abbe Center of Photonics, Friedrich Schiller University Jena, Max-Wien-Platz 1, 07743 Jena, Germany}
\affil[3]{Helmholtz	Institute Jena, Fröbelstieg 3, 07743 Jena, Germany}
\affil[4]{INRS-EMT, 1650 boulevard Lionel-Boulet, Varennes, Québec, Canada J3X1S2}
\affil[5]{Fritz Haber Institute of the Max Planck Society, 14195 Berlin, Germany}
\affil[6]{University of California at Berkeley, Department of Chemistry, Berkeley, CA 94720, USA}
\affil[7]{Fraunhofer Institute for Applied Optics and Precision Engineering IOF, Center for Excellence in Photonics, Albert-Einstein-Str. 7, 07745 Jena, Germany}

\affil[*]{Corresponding author: nils.geib@uni-jena.de}
\maketitle
\begin{abstract}
In this work, we demonstrate a discrete dispersion scan scheme using a low number of flat windows to vary the dispersion of laser pulses in discrete steps.
Monte Carlo simulations indicate that the pulse duration can be retrieved accurately with less than 10 dispersion steps, which we verify experimentally by measuring few-cycle pulses and material dispersion curves at 3 and 10 µm wavelength.
This minimal measuring scheme using only five optical components without the need for high-precision positioners and interferometric alignment can be readily implemented in many wavelength ranges and situations.
\end{abstract}

In the last decade, ultrafast laser sources in the mid-infrared~(MIR) have found many applications.
For example, femtosecond MIR pulses allow to probe the ultrafast vibration dynamics of molecules, which are of vital importance in physical, chemical, biological, and medical science~\cite{Lanin2014}.
Strong-field laser physics profits from the shift to longer wavelengths due to the scaling law of the ponderomotive energy $\propto \lambda^2$.
Here, using few-cycle MIR pulses allows for more efficient energy transfer from the laser field to the electrons~\cite{Wolter2015,Ghimire2019,Hollinger2020}.

To achieve few-cycle pulses in practice, a precise control and, therefore, characterization of the spectral phase of an octave-wide spectrum is required~\cite{Walmsley2009}.
One prominent method is frequency-resolved optical gating~(FROG)~\cite{Kane1993}.
In second-harmonic generation~(SHG) FROG the spectra of frequency-doubled pulses at the output of a non-collinear autocorrelator are measured for a range of delay positions.
The resulting two-dimensional measurement, the FROG trace, is a map of the SHG~intensity parametrized by wavelength and delay.
SHG-FROG and variants of FROG are used to measure pulses in a wide parameter range~\cite{Baltuska1999,Trebino2002}.
\begin{figure}[t]
\centering
\includegraphics[width=3.24in]{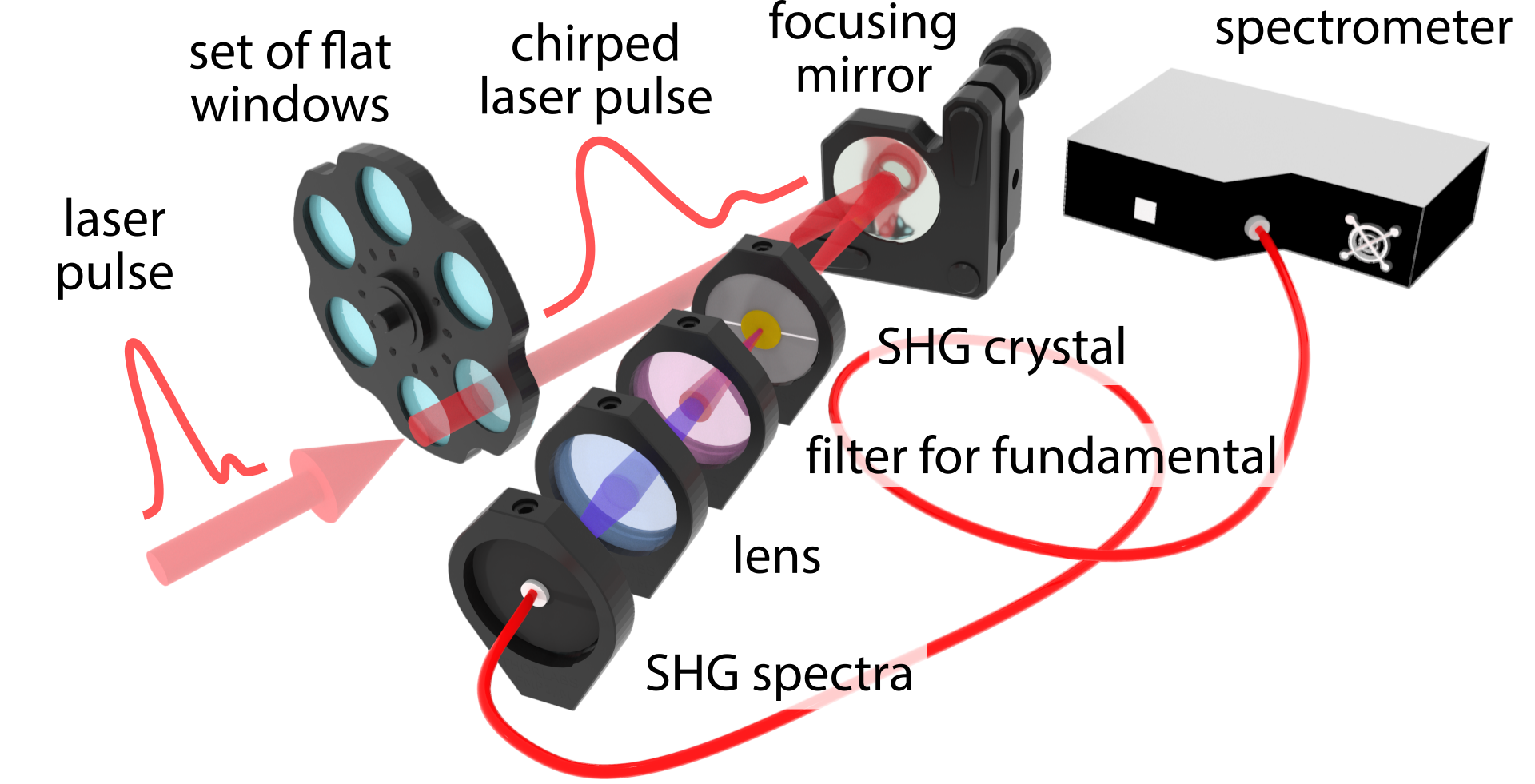}
\caption{A schematic of the simplified dispersion scan setup used for measuring few-cycle laser pulses at 3 and \SI{10}{\um}.}
\label{fig:setup}
\end{figure}

More recently, dispersion scan (d-scan) has emerged as a non-interferometric method allowing characterization of few- and single-cycle laser pulses~\cite{Miranda2012,Miranda2012a,Silva2014}.
In d-scan the pulse chirp is continuously varied by dispersive glass wedges and the second-harmonic spectra are measured.
The resulting two-dimensional measurement, the d-scan trace, is a map of the SHG intensity parametrized by wavelength and glass insertion~(dispersion).

D-scan is often integrated in a beamline for simultaneous characterization and compression of the pulses.
A pair of chirped mirrors or a grating compressor is used to introduce a negative chirp enabling effective net zero dispersion tuning by the glass wedges (positive chirp).
The wedge position for maximum temporal compression is then determined from the d-scan measurement.

FROG and d-scan both allow retrieving amplitude and phase of a pulse by means of a numerical algorithm.
Usually, a specialized algorithm based on generalized projections~\cite{DeLong1994,Kane2008} is used for FROG while general optimization algorithms are applied for d-scan~\cite{Miranda2012,Escoto2018}.
Recently, a common pulse retrieval algorithm~(COPRA) was proposed for both FROG and d-scan~\cite{Geib2019a}.

While FROG and d-scan were originally developed for the visible and near-infrared, they have been successfully transferred to the MIR~\cite{Bates2010,Bock2018,Silva2013}.
Also, recently, a technique based on transient absorption for pulse measurement up to \SI{10}{\um} has been demonstrated~\cite{Leblanc2019}.
In fact, the challenges in MIR pulse characterization are mainly of technical and not fundamental nature.
Detectors for measuring spectra beyond \SI{3}{\um} are single-pixel devices that require long exposures, severely impacting the measurement time.
Also, certain optical components may not be available or simply be costly.

Here we demonstrate a discrete and simplified d-scan scheme, that reduces the effort in setup and measurement~(see Fig.~\ref{fig:setup}). 
Compared to standard d-scan we use a small set of flat windows with different thicknesses as dispersive elements, removing the need for a pair of glass wedges and a calibrated high-precision stage.
We show experimentally that it is possible to retrieve pulse amplitude and phase with high accuracy with as little as seven discrete dispersion settings.
Our setup does not require a pre-compression, e.g., with a chirped mirror pair or a grating compressor, which could otherwise restrict the measurable bandwidth.
It is used for external pulse characterization complementary to in-line pulse compression with standard d-scan.

We want to point out one previous work which presents an experimentally similar technique under the same name~\cite{Wnuk2016}.
Here discrete amounts of positive dispersion are applied by increasing the number of passes through a flat window.
However, only the second and third order phase coefficients are retrieved from the peak positions of the measured SHG spectra.
No full amplitude and phase retrieval is performed, in contrast to this work.

The underlying question of our approach is how many dispersion steps are necessary for an accurate d-scan measurement.
The chirp-reversal technique~(CRT) demonstrated that two SH~spectra can uniquely determine pulse amplitude and phase~\cite{Loriot2013}.
Although it uses quadratic phase functions applied with a pulse shaper, this property is expected to be valid for d-scan at least approximately.
In contrast, d-scan is routinely performed with a large number of dispersion steps ($>100$) to sample the point of maximum pulse compression.
If used solely for pulse retrieval, far less measurements should be sufficient.

To study how the number of dispersion steps impacts the accuracy of the retrieved pulses, we performed Monte Carlo simulations.
We created $100$ random test pulses with time-bandwidth product of $2$ (see supplement of~\cite{Geib2019a}) and defined a fixed, positive dispersion range that was sampled with $M=2$ to $M=100$ steps.
For each configuration, we simulated and retrieved $10$ noisy d-scan traces and studied the ensemble of the solutions.

We looked at two metrics: first the pulse error, i.e., the normalized root-mean-square error~(NRMS) between the complex-valued spectrum of the retrieved and the test pulse, and second, the full width at half maximum~(FWHM) error, i.e., the relative difference between the FWHM pulse duration of the retrieved and the test pulse.
\begin{figure}[b!]
	\centering
	\includegraphics[width=3.2in]{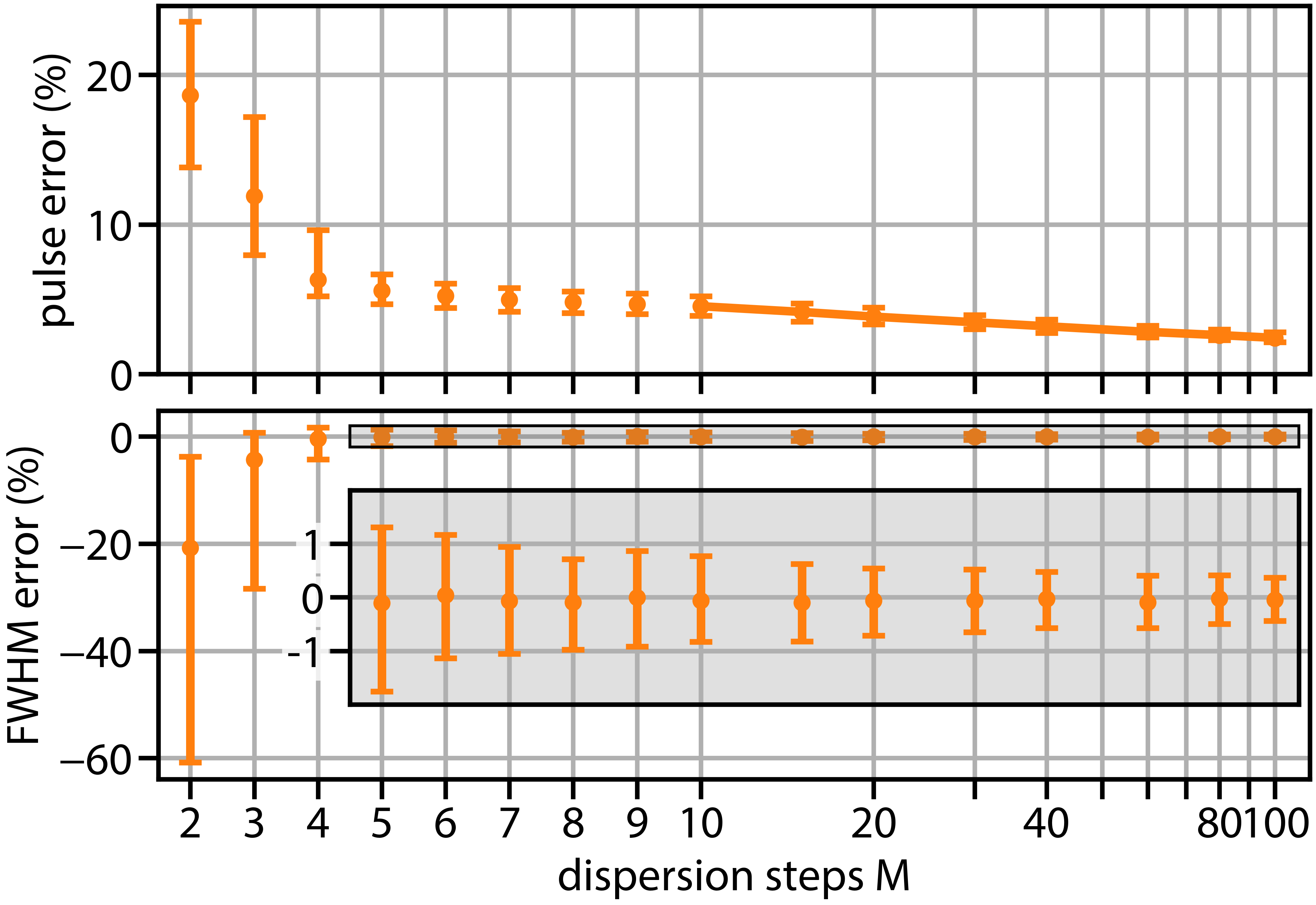}
	\caption{Accuracy of SHG-d-scan measurements in dependence of the number of discrete dispersion steps. Shown is the \SI{95}{\percent} prediction interval based on a Monte Carlo simulation of pulse retrieval from $1000$ measurements. For each data point, $10$ d-scan traces with Gaussian noise were generated for each of the $100$ random test pulses.}
	\label{fig:monte_carlo}
\end{figure}

We considered random measurement errors by adding Gaussian noise with the standard deviation $\sigma_{mn} = \rho_\mathrm{add} + \rho_\mathrm{mul} \T_{mn}$ to the d-scan trace $\T_{mn} \equiv \T(z_m, \omega_n)$ normalized to unity.
We chose the constant and intensity dependent components $\rho_\mathrm{add} = \SI{0.1}{\percent}$ and $\rho_\mathrm{mul} = \SI{3}{\percent}$ consistent with values observed in our SHG measurements at \SI{1.5}{\um}.
The pulses were retrieved by solving the least squares problem weighted by $\sigma_{mn}$ with COPRA.

The results are shown in Fig.~\ref{fig:monte_carlo}.
The mean and spread of the pulse errors reduces with the number of dispersion steps.
The curve has a large relative decrease for $M<5$ and moderate change for $M\geq5$.
An even stronger dependence can be seen in the FWHM error.
Here, the estimate is not reliable for $M \leq 3$, while for $M \geq 5$ the prediction interval for the FWHM error is below $\pm\SI{1.5}{\percent}$ for $M \geq 5$.
Furthermore, the benefit of increasing $M$ beyond $10$ is small.

We attribute the reduction in the errors mainly to the increase in data points, which reduces the influence of the measurement noise.
However, there seems to be a lower limit to the accuracy independent of $M$.
This may be attributed to the ill-conditioning of the retrieval problem, i.e., a nonlinear dependence of the pulse error on the trace error (the difference between measured and retrieved trace).
\begin{figure*}[t]
\centering
\includegraphics[width=7in]{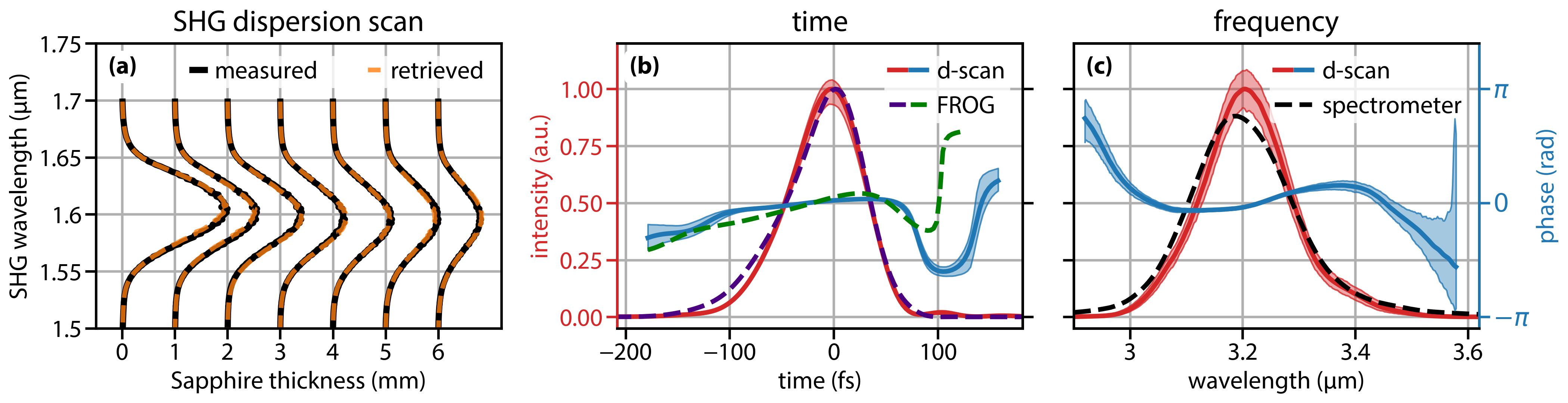}
\caption{Pulse at \SI{3}{\um} measured with the simplified dispersion scan setup: (a) the seven SHG spectra measured for different glass insertions (black) and the spectra calculated from the retrieved pulse (orange), (b) temporal intensity (red) and phase (blue) of the retrieved pulse compared to the same pulse measured with SHG-FROG (purple and green), (c) spectral intensity (red) and phase (blue) compared to the measured fundamental spectrum (black). The shaded regions are \SI{95}{\percent} confidence intervals estimated by bootstrap resampling.}
\label{fig:retrieval}
\end{figure*}
\begin{figure}[t]
\centering
\includegraphics[width=2.6in]{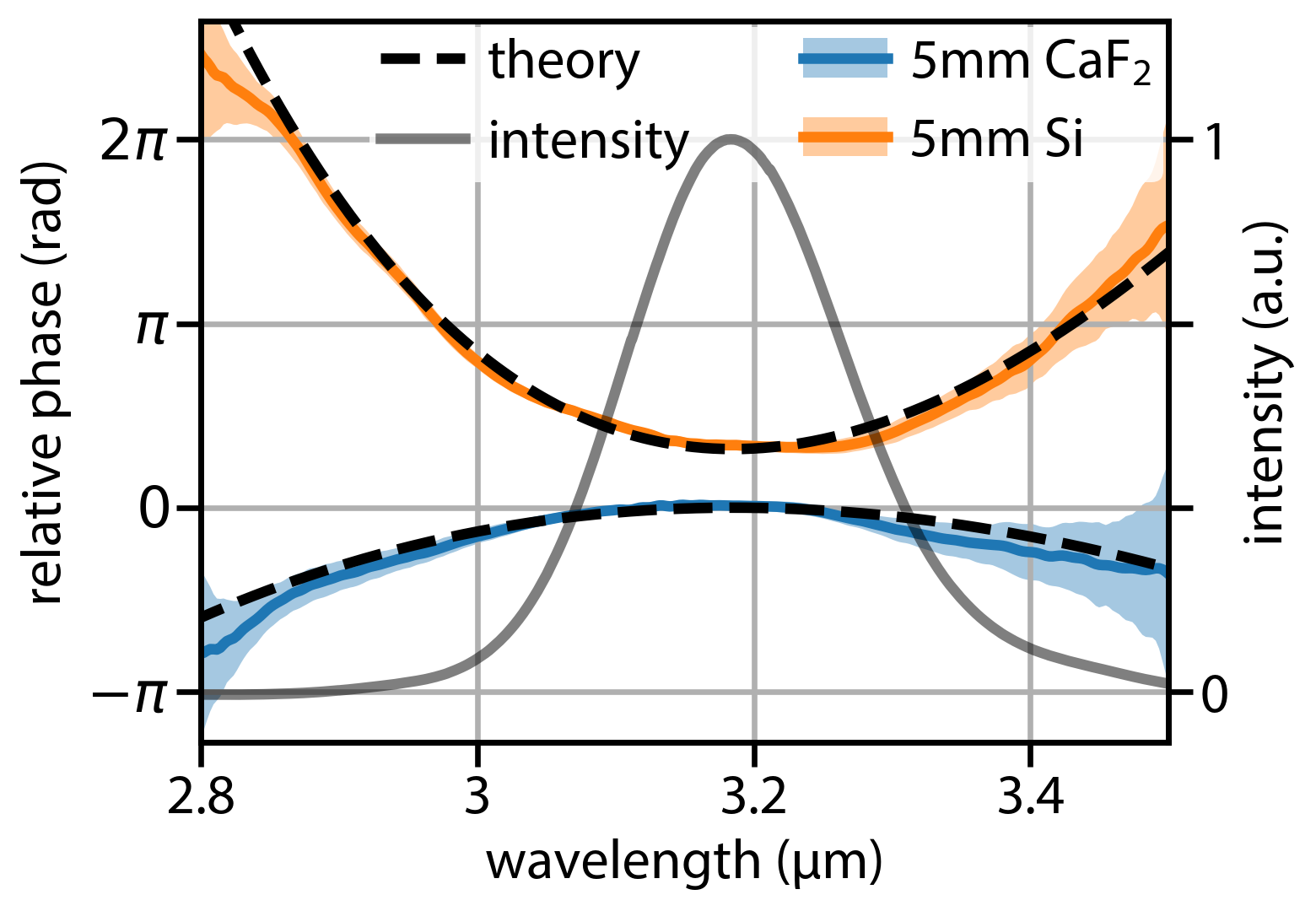}
\caption{Measurement of the relative material dispersion with our discrete d-scan setup: shown are the relative spectral phases introduced by $\mathrm{CaF}_\mathrm{2}$ (blue) or \SI{5}{\mm} of Si (orange) as obtained by measuring a pulse before and after propagation through the material. They are compared to the values calculated from Sellmeier equations of the materials (black). The shaded regions are \SI{95}{\percent} confidence intervals estimated by bootstrap resampling. The measured fundamental spectrum is shown in gray.}
\label{fig:dispersion}
\end{figure}
\begin{figure}[t]
\centering
\includegraphics[width=3.2in]{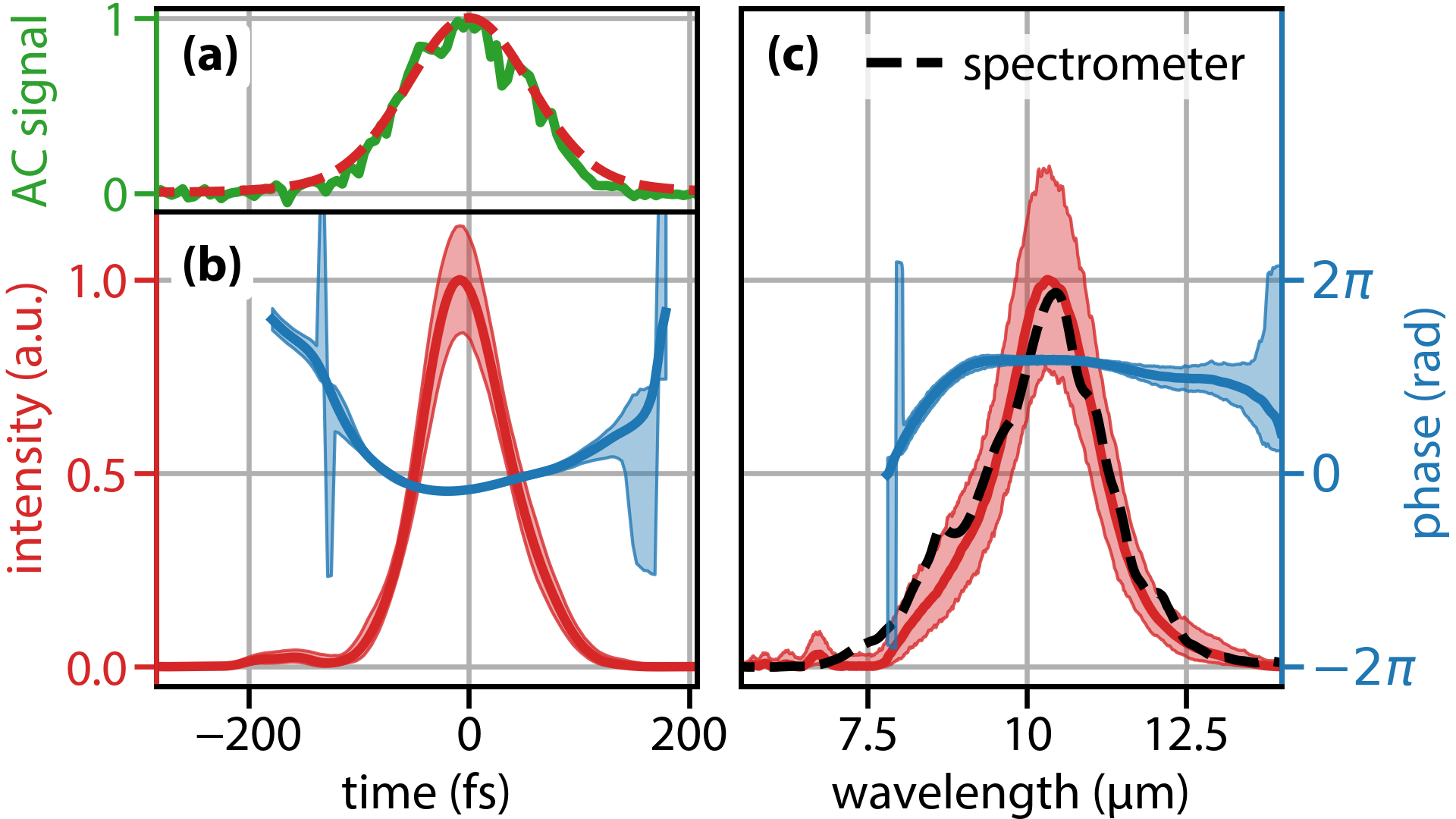}
\caption{Pulse at \SI{10}{\um} measured with the discrete dispersion scan setup: (a) measured (green) and retrieved (red) intensity autocorrelation (b) retrieved temporal intensity (red) and phase (blue), and (c) spectral intensity (red) and phase (blue) compared to the measured fundamental spectrum (black). The shaded regions are \SI{95}{\percent} confidence intervals estimated by bootstrap resampling.}
\label{fig:retrieval_10um}
\end{figure}

We conclude that using only a small amount of dispersion steps $M=5 \ldots 10$ has little drawbacks in terms of measurement accuracy, especially if one is mainly interested in the FWHM pulse duration.
We also observe that using $M=2$ or $M=3$ is not feasible for noise levels commonly encountered.
We also point out that often systematic measurement errors and calibration uncertainties rather than random noise will limit the retrieval accuracy in a d-scan measurement.
Due to their systematic nature we expect them to add a bias to the retrieval largely independent of the number of dispersion steps.
Thus, studying these effects is not necessary to support the conclusion above and beyond the scope of this work.

We proceeded to test our discrete d-scan setup experimentally.
First, we measured laser pulses at a central wavelength of \SI{3.2}{\um} with a pulse energy of \SI{30}{\micro\joule}.
The transform-limited pulse duration as calculated from the fundamental spectrum was \SI{61}{\fs} corresponding to $\num{5.7}$ optical cycles. 
The pulses were created by non-collinear difference frequency generation~(DFG) between the signal and idler waves from an optical parametric amplifier~(OPA) in a potassium titanyle arsenate~(KTA) crystal.
The OPA was pumped by \SI{5}{\milli\joule} pulses at \SI{800}{\nm}, with a repetition rate of \SI{1}{\kHz} and a duration of \SI{35}{\fs}.  

The dispersion was tuned by six sapphire windows with thicknesses between \SI{1}{\mm} and \SI{6}{\mm} in steps of \SI{1}{\mm}.
The windows were installed in a manual filter wheel, which was removed to obtain a seventh measurement at an insertion of \SI{0}{\mm}.
The number and thickness of the windows were chosen based on exploratory simulations to provide a high measurement accuracy.
Besides the number of windows, the maximum dispersion is important and has to be large compared to the measured pulse~\cite{Spangenberg2020}.

SHG was performed by focusing in a \SI{200}{\um} thick $\mathrm{AgGaS}_\mathrm{2}$~(AGS) crystal with a spherical mirror ($f = \SI{10}{\cm}$).
The frequency-doubled light was then imaged with a lens onto a multi-mode fiber coupled to an InGaAs array spectrometer.
For every dispersion step, 10~spectra were averaged and used to estimate the standard deviation of the measurement.
Besides the in-built dark current subtraction of the spectrometer no further post-processing, e.g., smoothing or apodization, was applied.
Pulse retrieval was performed using COPRA and followed by a final refinement with a general least squares solver~\cite{Geib2019a}.
The estimated standard deviation provided weights for the least squares problem.
The Fresnel reflections at the interfaces of the windows were included in the transfer function of d-scan to model the intensity loss compared to the spectrum obtained at \SI{0}{\mm} insertion.
The non-uniform spectral response of the measurement setup, e.g., phase matching, coupling efficiency, transmission of the fiber, and sensitivity of the spectrometer, were taken into account by retrieving a spectral response function simultaneously to the pulse~(see \cite{Miranda2017} or supplement of~\cite{Geib2019a}).

To estimate the uncertainty in the retrieved pulses from the measurement itself, we performed a bootstrap resampling.
Our procedure is similar to that in~\cite{Wang2003a} but uses the averaging from~\cite{Hause2015}.
We calculated \SI{95}{\percent} confidence intervals on intensity and phase from $\num{1000}$ bootstrap samples.
To estimate the actual accuracy of our method, we characterized the same laser with a self-built SHG-FROG setup.
It uses a pellicle beam splitter and the same SHG stage and spectrometer as the d-scan.
Pulses were retrieved using the same approach described above.
Finally, we measured the fundamental spectrum using an MIR spectrometer.

The results of the measurements are shown in Fig.~\ref{fig:retrieval}.
The pulse retrieved from the discrete d-scan measurement has a temporal duration of \SI{82.6(18)}{\fs}.
The pulse is weakly chirped compared to the transform limit of \SI{61}{\fs} and has a slight temporal asymmetry due to a cubic spectral phase. 

The retrieved fundamental spectrum agrees well with the measured one.
Especially, we point out how the tail of the spectrum above \SI{3.4}{\um} is retrieved correctly even though we did not measure any signal above \SI{1.7}{\um}, which was the limit of the spectrometer.
The reduced information for this wavelength range is reflected in the larger uncertainty in the spectral phase.
This apparent extrapolation is a known property of d-scan and similar schemes and is strictly physical because SHG is an auto-convolution in the spectral domain~\cite{Miranda2012a,Geib2019a}.

The pulse shape obtained from the SHG-FROG is in excellent agreement with the d-scan.
Its temporal duration of \SI{81.9(50)}{\fs} agrees within the confidence intervals.

To further analyse the accuracy of our method, we measured pulses before and after propagation through \SI{5}{\mm} of Si and $\mathrm{CaF}_\mathrm{2}$ with our d-scan setup~(see Fig.~\ref{fig:dispersion}).
From that we obtained the material dispersion without the constant and linear term in $k(\omega)$ as the relative spectral phase between both pulses.
This can be compared to Sellmeier values calculated for both materials.
The agreement with the theoretical values is excellent even at the edges of the spectrum where the intensity is low, albeit with a larger uncertainty at wavelengths for which no direct SHG was measured, i.e., above \SI{3.4}{\um}.

To demonstrate the feasibility of the method at longer wavelengths, we performed a second d-scan measurement of pulses at \SI{10}{\um} with \SI{40}{\micro\joule} pulse energy.
The transform-limited pulse duration as calculated from the fundamental spectrum was \SI{55}{\fs} corresponding to \num{1.6} optical cycles.
The pulses were generated by collinear DFG in GaSe between the signal and idler of a 3-stage OPA, pumped with \SI{1.1}{\milli\joule} pulses at \SI{800}{\nm}, with \SI{50}{\Hz} repetition rate and \SI{35}{\fs} duration.

Here, dispersion was tuned by a set of only five germanium windows, which were combined in pairs to form 8 dispersion settings with optical paths between \SI{3}{\mm} and \SI{11}{\mm}.
SHG was performed by focusing into a \SI{100}{\um} GaSe crystal with a parabolic mirror~($f = \SI{10}{\cm}$).
The SHG spectra were measured using a grating-based slit monochromator, with a cooled HgCdTe detector.

The retrieved pulse is shown in Fig.~\ref{fig:retrieval_10um}.
Its temporal duration is \SI{91(12)}{\fs}, consistent with the Gaussian pulse duration of \SI{92}{\fs} obtained from an independent autocorrelation measurement shown in Fig. \ref{fig:retrieval_10um}~(a).
The retrieved fundamental spectrum also shows good agreement with the measurement.
However, the relative uncertainty of the retrieved pulse duration and the spectrum is significantly larger than for the measurement at \SI{3}{\um}.
This is a consequence of the lower signal-to-noise ratio of the SHG spectra measured around \SI{5}{\um} that is propagated in the bootstrap estimation.

In summary, we have demonstrated that pulse characterization in amplitude and phase is possible with a discrete d-scan scheme that uses only a small set of flat windows as discrete dispersive elements.
Using Monte Carlo simulations, we analyzed the accuracy of d-scan in dependence on the number of dispersion steps. We could show that using $M = 5\ldots10$ steps is sufficient to measure the pulse duration with high accuracy.
Discrete d-scan was then applied to measure few-cycle pulses at 3 and \SI{10}{\um}.
The measurements at \SI{3}{\um} were verified by an independent SHG-FROG measurement and by comparing the retrieved dispersion of test materials to their known values.
The results showed that a discrete d-scan setup using only 7 and 8 steps for \SI{3}{\um} and \SI{10}{\um} few-cycle pulses, respectively, provides sufficient accuracy to compete with more complex measurement methods.

The demonstrated scheme takes the simplicity of d-scan to the next level.
It requires only five very basic optical components and omits high-precision positioners and the need for interferometric alignment.
While demonstrated in the MIR, there is no fundamental limitation in applying the scheme in the visible and near-infrared.
The minimal character of the setup makes it especially attractive in wavelength ranges or situations where components for more complex measurement setups are not available or cannot be used.
For example, in the MIR range, high-precision wedges for a standard continuous d-scan are expensive custom optics.

Finally, we want to point the reader's attention to the computer-driven approach adopted in this work.
Enabled by a fast and universal pulse retrieval algorithm~\cite{Geib2019a}, we could use Monte Carlo simulations to simplify standard d-scan while retaining most of the accuracy.
Similarly, the window number and thickness were selected based on simulations using expectation values for the pulses in the experiment.
This approach is beneficial for any pulse measurement scheme, e.g., the delay sampling in most FROG measurements could similarly be reduced by a large amount.
\medskip

\noindent\textbf{Acknowledgments.} NCG acknowledges funding by the German Research Council~(DFG) as part of CRC 1375 NOA and the German Federal Ministry of Education and Research~(BMBF) under grant~no.\ 03ZZ0467.
RH acknowledges financial support from the Helmholtz Institute Jena and the DFG for funding through the IRTG 2101.
FL and EH acknowledge support from NSERC, FRQNT, and PROMPT. 
EH acknowledges supports from NSERC Alexander Graham Bell fellowship.
MZ acknowledges funding by the BMBF under grant~no.\ 57427209, implemented by the German Academic Exchange Service (DAAD).
FE acknowledges funding by the BMBF and support by the Max Planck School of Photonics.
\medskip

\noindent\textbf{Disclosures.} The authors declare no conflicts of interest.

\bibliography{papers}

\begin{thebibliography}{10}

\bibitem{Lanin2014}
A.~Lanin, A.~Voronin, A.~Fedotov, and A.~Zheltikov, \enquote{Time-domain
  spectroscopy in the mid-infrared,} Sci. Rep. \textbf{4}, 6670 (2014).

\bibitem{Wolter2015}
B.~Wolter, M.~G. Pullen, M.~Baudisch, M.~Sclafani, M.~Hemmer, A.~Senftleben,
  C.~D. Schr{\"o}ter, J.~Ullrich, R.~Moshammer, and J.~Biegert,
  \enquote{Strong-field physics with mid-ir fields,} Phys. Rev. X \textbf{5},
  021034 (2015).

\bibitem{Ghimire2019}
S.~Ghimire and D.~A. Reis, \enquote{High-harmonic generation from solids,} Nat.
  Phys. \textbf{15}, 10--16 (2019).

\bibitem{Hollinger2020}
R.~Hollinger, D.~Hoff, P.~Wustelt, S.~Skruszewicz, Y.~Zhang, H.~Kang,
  D.~W\"{u}rzler, T.~Jungnickel, M.~Dumergue, A.~Nayak, R.~Flender, L.~Haizer,
  M.~Kurucz, B.~Kiss, S.~K\"{u}hn, E.~Cormier, C.~Spielmann, G.~G. Paulus,
  P.~Tzallas, and M.~K\"{u}bel, \enquote{Carrier-envelope-phase measurement of
  few-cycle mid-infrared laser pulses using high harmonic generation in zno,}
  Opt. Express \textbf{28}, 7314--7322 (2020).

\bibitem{Walmsley2009}
I.~A. Walmsley and C.~Dorrer, \enquote{Characterization of ultrashort
  electromagnetic pulses,} Adv. Opt. Photon. \textbf{1}, 308--437 (2009).

\bibitem{Kane1993}
D.~J. Kane and R.~Trebino, \enquote{Characterization of arbitrary femtosecond
  pulses using frequency-resolved optical gating,} IEEE J. Quant. Electron.
  \textbf{29}, 571--579 (1993).

\bibitem{Baltuska1999}
A.~Baltuska, M.~S. Pshenichnikov, and D.~A. Wiersma, \enquote{Second-harmonic
  generation frequency-resolved optical gating in the single-cycle regime,}
  IEEE J. Quantum Electron. \textbf{35}, 459--478 (1999).

\bibitem{Trebino2002}
R.~Trebino, \emph{Frequency-Resolved Optical Gating: The Measurement of
  Ultrashort Laser Pulses} (Springer US, 2000).

\bibitem{Miranda2012}
M.~Miranda, T.~Fordell, C.~Arnold, A.~L'Huillier, and H.~Crespo,
  \enquote{Simultaneous compression and characterization of ultrashort laser
  pulses using chirped mirrors and glass wedges,} Opt. Express \textbf{20},
  688--697 (2012).

\bibitem{Miranda2012a}
M.~Miranda, C.~L. Arnold, T.~Fordell, F.~Silva, B.~Alonso, R.~Weigand,
  A.~L'Huillier, and H.~Crespo, \enquote{Characterization of broadband
  few-cycle laser pulses with the d-scan technique,} Opt. Express \textbf{20},
  18732--18743 (2012).

\bibitem{Silva2014}
F.~Silva, M.~Miranda, B.~Alonso, J.~Rauschenberger, V.~Pervak, and H.~Crespo,
  \enquote{Simultaneous compression, characterization and phase stabilization
  of gw-level 1.4 cycle vis-nir femtosecond pulses using a single
  dispersion-scan setup,} Opt. Express \textbf{22}, 10181--10191 (2014).

\bibitem{DeLong1994}
K.~W. DeLong, B.~Kohler, K.~Wilson, D.~N. Fittinghoff, and R.~Trebino,
  \enquote{Pulse retrieval in frequency-resolved optical gating based on the
  method of generalized projections,} Opt. Lett. \textbf{19}, 2152--2154
  (1994).

\bibitem{Kane2008}
D.~J. Kane, \enquote{Principal components generalized projections: a review
  {[Invited]},} J. Opt. Soc. Am. B \textbf{25}, A120--A132 (2008).

\bibitem{Escoto2018}
E.~Escoto, A.~Tajalli, T.~Nagy, and G.~Steinmeyer, \enquote{Advanced phase
  retrieval for dispersion scan: a comparative study,} J. Opt. Soc. Am. B
  \textbf{35}, 8--19 (2018).

\bibitem{Geib2019a}
N.~C. Geib, M.~Zilk, T.~Pertsch, and F.~Eilenberger, \enquote{Common pulse
  retrieval algorithm: a fast and universal method to retrieve ultrashort
  pulses,} Optica \textbf{6}, 495--505 (2019).

\bibitem{Bates2010}
P.~K. Bates, O.~Chalus, and J.~Biegert, \enquote{Ultrashort pulse
  characterization in the mid-infrared,} Opt. Lett. \textbf{35}, 1377--1379
  (2010).

\bibitem{Bock2018}
M.~Bock, L.~von Grafenstein, U.~Griebner, and T.~Elsaesser, \enquote{Generation
  of millijoule few-cycle pulses at 5µm by indirect spectral shaping of the
  idler in an optical parametric chirped pulse amplifier,} J. Opt. Soc. Am. B
  \textbf{35}, C18--C24 (2018).

\bibitem{Silva2013}
F.~Silva, M.~Miranda, S.~Teichmann, M.~Baudisch, M.~Massicotte, F.~Koppens,
  J.~Biegert, and H.~Crespo, in \emph{CLEO: 2013,}  (Optical Society of
  America, 2013), p. CW1H.5.

\bibitem{Leblanc2019}
A.~Leblanc, P.~Lassonde, S.~Petit, J.-C. Delagnes, E.~Haddad, G.~Ernotte, M.~R.
  Bionta, V.~Gruson, B.~E. Schmidt, H.~Ibrahim, E.~Cormier, and
  F.~L\'{e}gar\'{e}, \enquote{Phase-matching-free pulse retrieval based on
  transient absorption in solids,} Opt. Express \textbf{27}, 28998--29015
  (2019).

\bibitem{Wnuk2016}
P.~Wnuk, H.~Fuest, M.~Neuhaus, L.~Loetscher, S.~Zherebtsov, E.~Riedle,
  Z.~Major, and M.~Kling, \enquote{Discrete dispersion scanning as a simple
  method for broadband femtosecond pulse characterization,} Opt. Express
  \textbf{24}, 18551--18558 (2016).

\bibitem{Loriot2013}
V.~Loriot, G.~Gitzinger, and N.~Forget, \enquote{Self-referenced
  characterization of femtosecond laser pulses by chirp scan,} Opt. Express
  \textbf{21}, 24879--24893 (2013).

\bibitem{Spangenberg2020}
D.-M. Spangenberg, E.~Rohwer, M.~Br\"{u}gmann, and T.~Feurer,
  \enquote{Extending time-domain ptychography to generalized phase-only
  transfer functions,} Opt. Lett. \textbf{45}, 300--303 (2020).

\bibitem{Miranda2017}
M.~Miranda, J.~Penedones, C.~Guo, A.~Harth, M.~Louisy, L.~Neori\v{c}i\'{c},
  A.~L'Huillier, and C.~L. Arnold, \enquote{Fast iterative retrieval algorithm
  for ultrashort pulse characterization using dispersion scans,} J. Opt. Soc.
  Am. B \textbf{34}, 190--197 (2017).

\bibitem{Wang2003a}
Z.~Wang, E.~Zeek, R.~Trebino, and P.~Kvam, \enquote{Determining error bars in
  measurements of ultrashort laser pulses,} J. Opt. Soc. Am. B \textbf{20},
  2400--2405 (2003).

\bibitem{Hause2015}
A.~Hause, S.~Kraft, P.~Rohrmann, and F.~Mitschke, \enquote{Reliable
  multiple-pulse reconstruction from second-harmonic-generation
  frequency-resolved optical gating spectrograms,} J. Opt. Soc. Am. B
  \textbf{32}, 868--877 (2015).

\end{thebibliography}

\end{document}